\documentclass[lettersize,journal]{IEEEtran}
\usepackage{cite}
\usepackage{stfloats}
\usepackage{amsmath,amsfonts,amssymb}
\usepackage{graphicx}
\usepackage{algorithm}
\usepackage{algorithmic}
\usepackage{epsfig}
\usepackage{float}
\usepackage{epstopdf}
\usepackage{hyperref}
\usepackage[caption=false]{subfig}
\usepackage{bm}
\usepackage{array}
\usepackage{bbm}
\usepackage{color}
\usepackage{textcomp}
\usepackage{verbatim}
\usepackage{balance}
\def\BibTeX{{\rm B\kern-.05em{\sc i\kern-.025em b}\kern-.08em
    T\kern-.1667em\lower.7ex\hbox{E}\kern-.125emX}}
\newtheorem{lemma}{Lemma}

\begin{document}
	\title{Optimizing Placement and Power Allocation in Reconfigurable Intelligent Sensing Surfaces for Enhanced Sensing and Communication Performance
	}
	\author{\IEEEauthorblockN{
			Cheng Luo, \IEEEmembership{Graduate Student Member, IEEE}, Jie Hu, \IEEEmembership{Senior Member, IEEE}, Luping Xiang, \IEEEmembership{Member, IEEE}, Kun Yang, \IEEEmembership{Fellow, IEEE} and Bo Lei
            }
			\\
        \thanks{Cheng Luo, Jie Hu, and Luping Xiang are with the School of Information and Communication Engineering, University of Electronic Science and Technology of China, Chengdu, 611731, China, email: chengluo@std.uestc.edu.cn; hujie@uestc.edu.cn; luping.xiang@uestc.edu.cn.}
        \thanks{Kun Yang is with the School of Computer Science and Electronic Engineering, University of Essex, Colchester CO4 3SQ, U.K., email: kunyang@essex.ac.uk}
		\thanks{Bo Lei is with the Research Institute of China Telecom Co., Ltd., Beijing, 102209, China, email: leibo@chinatelecom.cn}
	}
	\maketitle

	\begin{abstract}
		In this letter, we investigate the design of multiple reconfigurable intelligent sensing surfaces (RISSs) that enhance both communication and sensing tasks. An RISS incorporates additional active elements tailored to improve sensing accuracy. Our initial task involves optimizing placement of RISSs to mitigate signal interference. Subsequently, we establish power allocation schemes for sensing and communication within the system. Our final consideration involves examining how sensing results can be utilized to enhance communication, alongside an evaluation of communication performance under the impact of sensing inaccuracies. Numerical results reveal that the sensing task reaches its optimal performance with a finite number of RISSs, while the communication task exhibits enhanced performance with an increasing number of RISSs. Additionally, we identify an optimal communication spot under user movement.
	\end{abstract}


\section{Introduction}

\IEEEPARstart{I}{ntegrated} Sensing And Communication (ISAC) has emerged as a promising technology for the upcoming sixth-generation (6G) networks, attracting extensive research efforts across various domains, including ISAC framework development and power allocation strategies\cite{ISAC_powerallocation}, simultaneous secure communication and sensing\cite{ISAC_commandsensing, xuke_IoT}, and ISAC waveform design\cite{ISAC_waveformdesign}. Concurrently, Reconfigurable Intelligent Surfaces (RISs) have also gained significant attention as a key enabler for future wireless networks, with potential applications in autonomous driving, smart factories, and low-latency/low-power systems\cite{IRS_Tutorial}. The integration of RIS with ISAC offers further enhancements due to RIS's cost-effectiveness, ease of deployment, and ability to improve coverage, which extends the technology's applicability to innovative areas such as Channel State Information (CSI) acquisition\cite{IRS_CSI}, beam tracking\cite{IRSbeamtracking}, and echo enhancement\cite{echo_enhancement1}.

Recently, significant attention has been directed toward a unique RIS configuration that integrates both active and passive elements \cite{RISS_4, wqq_RISS,JSAC_RISS, selfsensingIRS}, hereinafter referred to as reconfigurable intelligent sensing surface (RISS). The incorporation of active elements within the RISS augments its sensing capabilities and significantly mitigates path loss caused by echo signals during sensing tasks. Conventionally, sensing signals experience a four-phase path loss, encompassing BS-RIS, RIS-target, the reverse target-RIS, and RIS-BS paths. However, this path is simplified to BS-RISS, RISS-target and target-RISS with the assistance of the RISS. Moreover, processing the echo signal before its reflection to the BS aids maintain the signal's spatial characteristic, since the signals consistently reflecting from the RIS hinder the identification of target angle information at the BS. Specifically, \cite{RISS_4} presents a multi-RISS scenario where each RISS supports both sensing and communication by optimizing beamformers to reduce sensing errors while ensuring high communication quality. \cite{wqq_RISS} introduced a joint ToA and DoA estimation method for localization with RISS assistance, analyzing the Cramér-Rao Bound (CRB) and proposing a low-complexity, non-iterative algorithm. In \cite{JSAC_RISS}, a sensing signal transmitted from the RISS controller and reflected by the RISS is used to enhance sensing effectiveness. Additionally, \cite{selfsensingIRS} highlights the importance of RISS in assisting communication within wirelessly powered networks.

Despite considerable research into RIS-based sensing and communication, most studies have primarily focused on the sensing capabilities of a single RIS and the related design of sensing algorithms for RIS-oriented systems. However, there is still a gap in designing sensing and communication schemes for multi-RIS systems and analyzing the impact of sensing errors on communication performance.

In light of the presented background, our study primarily focuses on the design of sensing and communication power allocations for multi-RISSs systems and the analysis of the performance of sensing-assisted communication under imperfect sensing. Our contributions can be summarized as follows:
\begin{itemize}
    \item We proposed a system with the aid of multiple RISSs. The RISSs attenuate the path loss of echo signals for sensing tasks, and enhance the communication effectiveness. This system holds significant implications for the future implementation of RISs in sensing and communication.
    \item We first formulate the placement strategies for multiple RISSs with an aim to neutralize signal interference. Given the placement locations, we implement different power allocation schemes for the sensing scheme and communication scheme to achieve optimal performance. Moreover, numerical results illustrate the impact of the number of RISSs on communication and sensing performance.
    \item We further scrutinize the communication performance by utilizing imperfect sensing results (i.e., imperfect DOA information) as an alternative for CSI estimation. Numerical results proved the correctness of our analysis.
\end{itemize}

\begin{figure}
	\centering
	\includegraphics[width=0.9\linewidth]{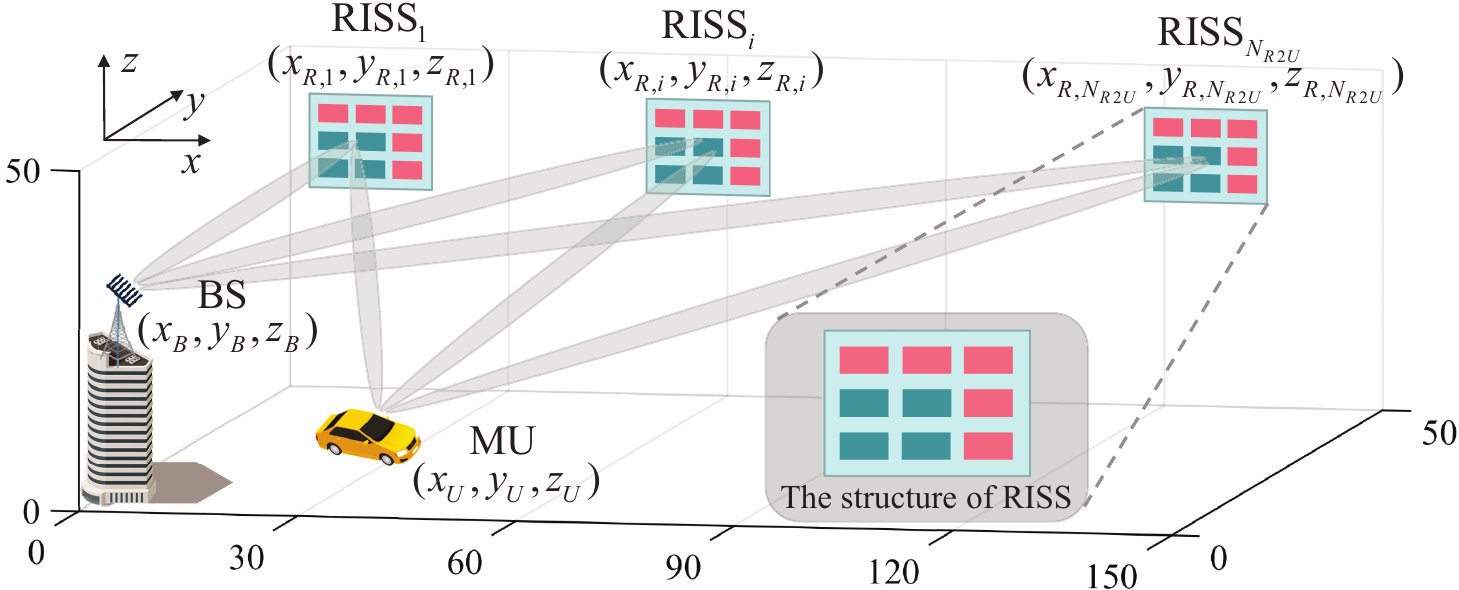}
	\setlength{\abovecaptionskip}{0pt}
    \setlength{\belowcaptionskip}{0pt} 
	\caption{A multiple RISSs-assisted sensing and communication system.}
	\label{fig:systemmodel}
\end{figure}

\section{System Model}
We consider a multiple RISSs assisted sensing and communication system as depicts in Fig. \ref{fig:systemmodel}, which comprises a Base Station (BS) with $M$ antennas and $N_{R2U}$ RISSs. Each RIS is composed of $N=N_x\times N_y$ passive elements and $N_a$ active elements, enabling sensing and communication simultaneously. The system assumes the presence of a single Mobile User (MU) with a single antenna. Let $\mathbf{s} = [s_1, \cdots, s_{N_{R2U}}]^T \in \mathbb{C}^{N_{R2U} \times 1}$ represent the Gaussian data symbols transmitted from the BS to the RISSs, where $\mathbb{E}\{\mathbf{s}\mathbf{s}^H\} = \mathbf{I}$.

Let $\mathbf{G}_k\in\mathbb{C}^{N\times M}$ and $\mathbf{h}_k\in\mathbb{C}^{N\times 1 }$ denote the channel from the BS to $k$-th RISS and  $k$-th RISS to the MU, we have
\begin{align}
	&\mathbf{G}_k = \boldsymbol{\alpha}(\vartheta_{G,k}, \varphi_{G,k})\boldsymbol{\beta}^T(\varpi_{G,k}), \\
	&\mathbf{h}_{k} = \boldsymbol{\alpha}(\vartheta_{h,k}, \varphi_{h,k}),
\end{align}
where $\left[\boldsymbol{\alpha}(\vartheta, \varphi)\right]_n =  e^{(\text{mod}(n,N_y)-1)\mathbbm{i}\vartheta}e^{(\lfloor n/N_x+1 \rfloor-1)\mathbbm{i}\varphi}$, $\left[\boldsymbol{\beta}(\varpi)\right]_m = e^{(m-1)\mathbbm{i}\varpi}$ and $\varphi=2\pi d\cos(\phi_{\text{azi}})/\lambda=\pi\cos(\phi_{\text{azi}})$ by setting $d/\lambda=1/2$ without loss of generality, with $d$ and $\lambda$ are the element spacing and carrier wavelength. Similarly, we have $\vartheta = \pi\sin(\phi_{\text{azi}})\sin(\phi_{\text{ele}})$, $\varpi=\pi\sin(\phi_{\text{dep}})$. And $\phi_\text{azi}$, $\phi_\text{ele}$ and $\phi_\text{dep}$ denote the angle of arrival (AoA) and angle of departure (AoD), respectively.

With the assistance of $N_{R2U}$ RISSs, we can obtain the downlink signals received by MU as
\begin{align}
	y_{MU}=\sum_{k=1}^{N_{R2U}}\sum_{i=1}^{N_{R2U}}\varrho_{B2R,k}\varrho_{R2U,k}\mathbf{h}_k^T\boldsymbol{\Theta}_k\mathbf{G}_k\mathbf{w}_is_i+n,\label{eqn:ymu}
\end{align}
where $\varrho_{B2R,k}=\sqrt{\frac{\lambda^2}{16\pi^2 d_{B2R,k}^2}}$ and $\varrho_{R2U,k}=\sqrt{\frac{\lambda^2}{16\pi^2 d_{R2U,k}^2}}$ denote the pathloss from the BS to the $k$-th RISS and the $k$-th RISS to the MU. These pathloss are determined based on the distances between the BS and the $k$-th RISS, denoted as $d_{B2R,k}$, and between the $k$-th RISS and the MU, denoted as $d_{R2U,k}$. The additive Gaussian noise, represented as $n \sim \mathcal{CN}(0, \sigma_0^2)$, follows a complex Gaussian distribution with zero mean and noise power $\sigma_0^2$. $\mathbf{w}_k$ and $\boldsymbol{\Theta}_k$ denote the precoder and the diagonal phase shift matrix of the $k$-th RISS, respectively.

In addition, we assume that the active elements of each RISS are arranged in an L-array. These active elements can acquire echo signals to estimate the AoA of the MU during the sensing scheme, which in turn aids in the passive phase shift design for the communication scheme. Given an incident signal $y_{\text{MU}}$, the corresponding echo signal received by the active elements of the $k$-th RISS can be expressed as
\begin{align}
	\mathbf{y}_{RISS,k} = [\varrho_{U2R,k}\boldsymbol{\beta}(\vartheta_{r,k}) y_{MU};\varrho_{U2R,k}\boldsymbol{\beta}(\varphi_{r,k}) y_{MU}],
\end{align}
where $\varrho_{U2R,k}=\sqrt{\frac{\varsigma}{4\pi d_{U2R,k}^2}}$, $d_{U2R,k}$ and $\varsigma$ denote the distance between the MU to the $k$-th RISS and Radar Cross-Section (RCS), respectively. 
\begin{figure}
	\centering
	\includegraphics[width=0.72\linewidth]{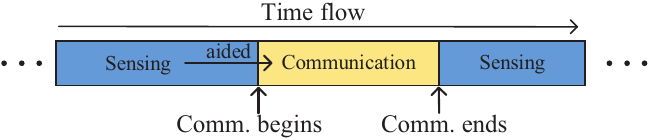}
	\setlength{\abovecaptionskip}{0pt}
    \setlength{\belowcaptionskip}{0pt} 
	\caption{The frame structures of sensing and communication.}
	\label{fig:frame_structure}
\end{figure}

Fig. \ref{fig:frame_structure} illustrates the frame structures of the sensing and communication schemes and their relationship. Both schemes share the same hardware. The system primarily operates in sensing scheme but switches to communication scheme when a MU enters and communication begins, leveraging sensing information (e.g., DOA information) as a substitute for CSI.

\section{The Proposal of Multiple RISSs Systems}
For the purpose of interference elimination, we initially design the placements of the multiple RISSs. Subsequently, based on the placement, we then develop two operational schemes for the sensing and communication tasks. Finally, we conducted a performance analysis of above systems\footnote{The proposed system can be extended to a MIMO system, as the echo signal power in the sensing scheme is independent of the number of antennas at the MU, and the received power in the communication scheme can be enhanced using combiners like maximal ratio combining (MRC). However, multi-user interference must be addressed for scenarios with multiple users, which we will explore in future work.}.
% Note that the proposed multiple-input-single-output (MISO) system can be directly extended to a multiple-input-multiple-output (MIMO) system. This is because the echo signal power in the sensing scheme is independent of the number of antennas at the MU, and the received power in the communication scheme can be further enhanced using an appropriate combiner, such as maximal ratio combining (MRC). However, it is necessary to consider multi-users interference elimination for multi-user scenario. We will explore these aspects in our future work.

\subsection{Placement Design of Multiple RISSs}\label{sec:A}
To mitigate interference among signals transmitted by the BS to various RISSs and reduce the complexity of coupled passive phase shift design of each RISS and precoder of the BS, we perform low-complexity closed-form solutions. Specifically, we expand Eq.\eqref{eqn:ymu} further and obtain
\begin{align}
	&y_{MU}=\sum_{k=1}^{N_{R2U}}\varrho_{B2R,k}\varrho_{R2U,k}\mathbf{h}_k^T\boldsymbol{\Theta}_k\mathbf{G}_k\sum_{i=1}^{N_{R2U}}\mathbf{w}_is_i\nonumber\\
	&=\sum_{k=1}^{N_{R2U}}\varrho_{B2R,k}\varrho_{R2U,k}\mathbf{h}_k^T\boldsymbol{\Theta}_k\mathbf{G}_k\bigg(\mathbf{w}_k+\underbrace{\sum_{i\neq k}^{N_{R2U}}\mathbf{w}_is_i}_{\text{leakage signals}}\bigg).
\end{align}
\begin{lemma} \label{lemma:lemma1}
	When the leakage signals are eliminated, an optimal decoupling solution exists for $\boldsymbol{\Theta}_k$ and $\mathbf{w}_k$ as
	\begin{align}
		&\boldsymbol{\Theta}_k=\text{diag}\left\{\left(\boldsymbol{\alpha}(\vartheta_{h,k}, \varphi_{h,k})\circ\boldsymbol{\alpha}(\vartheta_{G,k}, \varphi_{G,k})\right)^\dagger\right\},\label{eqn:best_theta}\\
		&\bar{\mathbf{w}}_k=\sqrt{\eta_k}\frac{\beta^\dagger(\varpi_{G,k})}{||\beta(\varpi_{G,k})||},\label{eqn:bestw}
	\end{align}
	where $\circ$ and $(\cdot)^\dagger$ denote the Hadamard product and conjugate operator, and $\eta_k$ denote the transmit power for $k$-th RISS. And the interference signals can be effectively eliminated with $\left|\sin(\varpi_{G,k})-\sin(\varpi_{G,i})\right|=\frac{2n}{M},n=1,2,\cdots.$
\end{lemma}
\begin{IEEEproof}
	Please refer to Appendix \ref{app:A} for detailed proof.
\end{IEEEproof}
% Due to the space limitation, the proof is omitted here and is shown in a longer version of this paper \cite{xuke_IoT}.
Let $(x_B, y_B, z_B)$, $(x_U, y_U, z_U)$ and $(x_{R,k}, y_{R,k}, z_{R,k})$ denote the locations of the BS, MU, and the $k$-th RISS, respectively. For simplicity, all RISSs and the BS are positioned at the same elevation, i.e., $z_B=z_{R,k},\forall k\in{N_{R2U}} $. Additionally, the RISSs are arranged in a linear configuration along the y-axis, such that $y_{R,j}=y_{R,k}=R_r,\forall k,j\in {N_{R2U}}$ represents the distance from the BS to any RISS along the y-axis. The optional placement of the RISSs, denoted by $\{x_{R,l}\}_{l=1}^L$, is designed to satisfy the orthogonal angle condition. This condition is expressed as follows:
\begin{align}
	\{x_{R,l}\}_{l=1}^L = \frac{2lR_r}{\sqrt{M^2-4l^2}}, l=1,\cdots, L,M\geq2L.\label{eqn:quantloc}
\end{align}

Eq. \eqref{eqn:quantloc} defines the final optional placement configuration for the RISSs, ensuring that they are positioned at orthogonal angles to effectively eliminate leakage signals.

Since MU may enter the service area of sensing and communication from various directions, it is necessary to consider distributing RISSs as uniformly as possible in space. Moreover, in order to fulfill the aforementioned orthogonality requirement, the quantized positions resulting from the uniform distribution need to be incorporated into Eq. \eqref{eqn:quantloc}. In this letter, we uniformly distribute and quantify the locations of RISS between $[0, \frac{2(L-1)R_r}{\sqrt{M^2-4(L-1)^2}}]$. 

\subsection{Multiple RISSs assisted Sensing Scheme Design}
Before delving into the design of the sensing scheme, we present a concise definition of the sensing detectable range. The detectable range of the $k$-th RISS is defined as the maximum extent within which the echo signal, received at each active element of the RISS, exceeds a predefined signal-to-noise ratio (SNR) threshold denoted as $\Gamma$. This implies that the sensing detectable range of the RISS can be expressed as
\begin{align}
	\frac{\left|[\mathbf{y}_{RISS,k}]_i\right|^2}{\sigma_0^2}\geq \Gamma, i\in N_a.
\end{align}
Further expansion yields
\begin{align}
	d_{R2U,k}\leq\sqrt[4]{\frac{\lambda^2\varsigma\left|\varrho_{B2R,k}\mathbf{h}_k^T\boldsymbol{\Theta}_k\mathbf{G}_k\mathbf{w}_ks_k\right|^2}{64\pi^3\Gamma\sigma_0^2}}.	
\end{align}

Let $\mathcal{A}_{\cup}$ denotes the total detectable range of $N_{R2U}$ RISSs, which can be expressed as
\begin{align}
	\mathcal{A}_{\cup} = \bigcup_{k\in N_{R2U}}\mathcal{A}_k(d_{R2U,k}),
\end{align}
where $\mathcal{A}_k(d_{R2U,k})$ represents the detectable range of the $k$-th RIS. To effectively handle the sensing task for an unknown MU, it is essential that different RISSs have as uniform a detectable range as possible. This uniformity is achieved by allocating more power to RISSs located farther from the BS and less to those that are closer. The power allocation strategy is therefore formulated as a max-min optimization problem, resulting in a solution that balances the power distribution across the RISSs as
\begin{align}
	&\text{(P1): }\max_{\mathbf{w}_k,\boldsymbol{\Theta}_k} \min_{k\in N_{R2U}}\left|\varrho_{B2U,k}\varrho_{U2R,k}\mathbf{h}_k^T\boldsymbol{\Theta}_k\mathbf{G}_k\mathbf{w}_ks_k\right|^2\label{eqn:max_min1}\\
	&\qquad \quad \text{s.t. } \, \sum_{i=1}^{N_{R2U}}||\mathbf{w}_k||^2\leq P,\tag{\ref{eqn:max_min1}a}\label{eqn:max_min1a}\\
	&\qquad \qquad\quad |[\boldsymbol{\Theta}]_{i,i}|=1,\forall i\in N,\tag{\ref{eqn:max_min1}b}\label{eqn:max_min1b}
\end{align}

where $\varrho_{B2U,k} = \varrho_{B2R,k}\varrho_{R2U,k}$. Objective $\text{(P1)}$ represents maximizing the minimal power of RISSs received. $\mathbf{w}_k$ and $\boldsymbol{\Theta}_k$ are the transmit precoder of the BS to the $k$-th RISS and the phase shift matrix of the RISS. Total power denotes by $P$ and passive phase shift limits cause constraints in \eqref{eqn:max_min1a} and \eqref{eqn:max_min1b}, respectively.

By substituting Eq. \eqref{eqn:bestw} and Eq. \eqref{eqn:best_theta} \footnote{We assume the knowledge of $\vartheta_{G,k}$ and $\varphi_{G,k}$, which can be obtained from the placement of the RISSs or active elements' DoA estimation. Additionally, $\vartheta_{h,k}$ and $\varphi_{h,k}$ are considered fixed when maximizing the echo signal power.} into problem $(\text{P1})$, we have $\mathbf{h}_k^T\mathbf{\Theta}_k\mathbf{G}_k\mathbf{w}_k=\boldsymbol{\alpha}(\vartheta_{h,k}, \varphi_{h,k})^T\mathbf{\Theta}_k\boldsymbol{\alpha}(\vartheta_{G,k}, \varphi_{G,k})\sqrt{\eta_k}\sqrt{M}=N\sqrt{M}\sqrt{\eta_k}$. Thus, $\text{(P1)}$ can be further simplified as
\begin{align}
	&\text{(P1.1): }\max_{\eta_k,k\in N_{R2U}}\,\, t \label{eqn:max_min12}\\
	&\qquad \qquad \,\,   \text{s.t. } \sum_{i=1}^{N_{R2U}}\eta_i\leq P,\tag{\ref{eqn:max_min12}a}\label{eqn:max_min12a}\\
	&\qquad \qquad\,\,\qquad \varrho_{B2R,i}^2\eta_i\geq t, \forall i\in N_{R2U}\tag{\ref{eqn:max_min12}b},\label{eqn:max_min12b}
\end{align}

where $t$ is an auxiliary variable, and $\eta_i,\forall i\in N_{R2U}$ represents the transmit power for the $i$-th RISS. It is important to note that $\text{(P1.1)}$ is a convex optimization problem, which can be efficiently solved using standard optimization techniques, such as the interior-point method. The complexity of solving $\text{(P1.1)}$ using the interior-point method is $\mathcal{O}(N_{R2U}^{3.5})$.

\subsection{Multiple RISSs assisted Communication Scheme Design}
After obtaining sensing information of a specific MU during the sensing phase, the entire system transitions from the sensing scheme to the communication scheme under the invariant placements of RISSs. The sensing results are then utilized as a substitute for channel estimation and aid in the communication scheme. In this section, we initially assume perfect sensing angle information to devise the power allocation scheme.

Assuming that the BS transmits the same signal, denoted as $s$, to different RISSs for communication tasks, the received signal at the MU varies due to the distance from the BS to each RISS, yielding
\begin{align}
	y_{MU}=\sum_{k=1}^{N_{R2U}}\sum_{i=1}^{N_{R2U}}\varrho_{B2U,k}\mathbf{h}_k^T\boldsymbol{\Theta}_k\mathbf{G}_k\mathbf{w}_is_ie^{-\mathbbm{i}\Delta\phi_k}+n,
\end{align}
where the time delay from the BS to the MU can be ignored, as the system is considered narrowband, and $e^{-\mathbbm{i}\Delta\phi_k}$ denotes the phase of the corresponding path. We define $\boldsymbol{\Theta}^\S_k=\boldsymbol{\Theta}_k\otimes e^{\mathbbm{i}\Delta\phi_k}$ for subsequent investigations. Based on Section \ref{sec:A}, we have
\begin{align}
	&\text{(P2):}\max_{\substack{\mathbf{w}_k,\boldsymbol{\Theta}^\S_k}}\log_2\left(1+\left|\sum\limits_{k=1}^{N_{R2U}}\varrho_{B2U,k}\mathbf{h}_k^T\boldsymbol{\Theta}^\S_k\mathbf{G}_k\mathbf{w}_k\right|^2\middle/\sigma^2_0\right)\label{eqn:maxsumrate1}\\
	&\qquad \quad \text{s.t. } (\ref{eqn:max_min1}a), (\ref{eqn:max_min1}b),\tag{\ref{eqn:maxsumrate1}a}\label{eqn:maxsumrate1a}
\end{align}
where the objective $\text{(P2)}$ represents maximizing weighted sum channel capacity. By substituting Eq. \eqref{eqn:bestw} and Eq. \eqref{eqn:best_theta} into objective $(\text{P2})$, and due to the monotonicity of function $\log_2(\cdot)$, Eq. \eqref{eqn:maxsumrate1} is equivalent to
\begin{align}
	&\text{(P2.1): }\max_{\eta_k,k\in N_{R2U}} \sum_{k=1}^{N_{R2U}}\varrho_{B2U,k}N\sqrt{M}\sqrt{\eta_k}\label{eqn:maxsumrate11}\\
	&\qquad\qquad\, \text{s.t. } (\ref{eqn:max_min12}a),\tag{\ref{eqn:maxsumrate11}a}\label{eqn:maxsumrate11a}
\end{align}
where $\text{(P2.1)}$ represents the objective of maximizing the weighted sum power received by the MU with the assistance of $N_{R2U}$ RISSs. Problem $\text{(P2.1)}$ remains convex and can be addressed using standard optimization techniques, such as the interior-point method. The computational complexity of solving $\text{(P2.1)}$ using this approach is also $\mathcal{O}(N_{R2U}^{3.5})$.

Note that the communication and sensing schemes share the same system components, such as the BS and the $N_{R2U}$ RISSs, but have different power allocation due to their distinct objectives. The sensing scheme provides DOA information, replacing CSI and supporting beamforming in subsequent communication. 
\begin{figure}[t]
	\centering
	\includegraphics[width=0.65\linewidth]{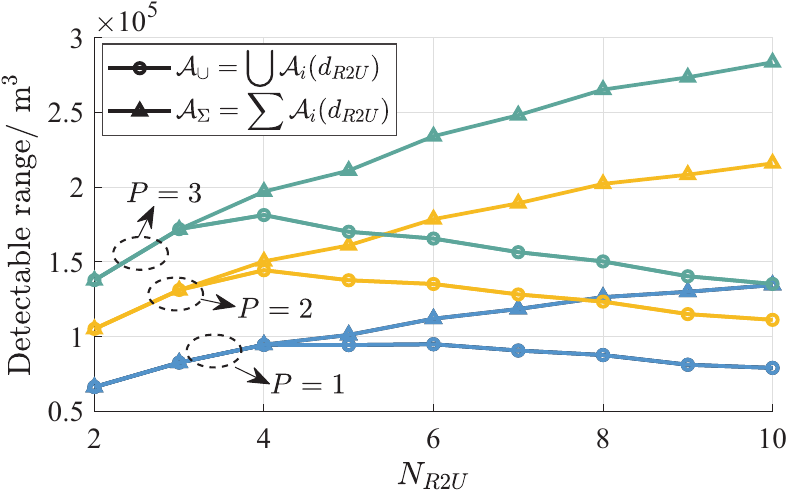}
	\setlength{\abovecaptionskip}{0pt}
    \setlength{\belowcaptionskip}{0pt} 
	\caption{The detectable range varies with $N_{R2U}$ and the transmit power $P$.}
	\label{fig:exp1sensingarea}
\end{figure}
\begin{figure}[t]
	\centering
	\includegraphics[width=0.65\linewidth]{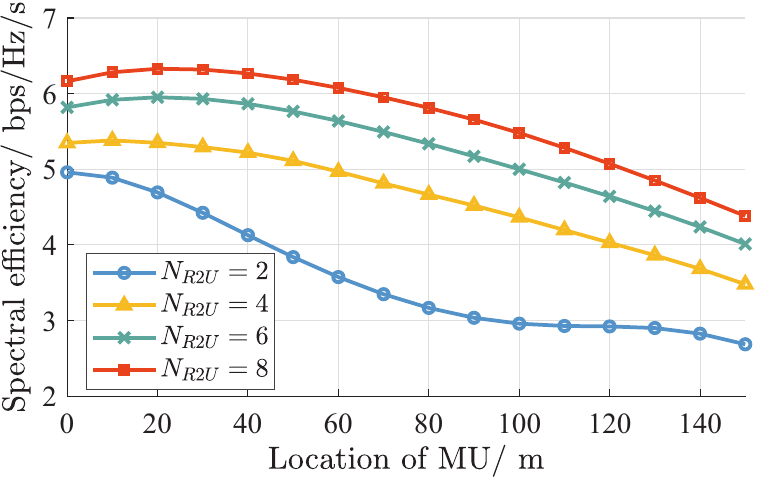}
	\setlength{\abovecaptionskip}{0pt}
    \setlength{\belowcaptionskip}{0pt} 
	\caption{The spectral efficiency varies with $N_{R2U}$ and the location of the MU.}
	\label{fig:exp2spectral_effi}
\end{figure}
\subsection{Performance Analysis of Sensing assisted Communication}\label{sec:imperfectsensing}
After introducing the design of the sensing and communication schemes, we now consider the impact of imperfect sensing on the communication performance in our model. Assuming that we employ the Root Multiple Signal Classification (ROOT-MUSIC) algorithm in our sensing scheme, thus, we can model the sensing error of the array response vector as a Gaussian distribution. Specifically, we can express it as
\begin{align}
	&\varphi_{t,k}^e-\varphi_{t,k}=\xi_{t,\varphi,k}\sim\mathcal{N}(0, \sigma_{t,\varphi,k}^2), t\in\{h, G\}\nonumber\\
	&\vartheta_{t,k}^e-\vartheta_{t,k}=\xi_{t,\vartheta,k}\sim\mathcal{N}(0, \sigma_{t,\vartheta,k}^2), t\in\{h, G\},
\end{align}
where $\varphi_{h,k}^e$, $\vartheta_{h,k}^e$, $\varphi_{G,k}^e$, $\vartheta_{G,k}^e$ represent the array response vectors obtained from the sensing with errors (i.e., $\xi_{h,\varphi,k}$, $\xi_{h,\vartheta,k}$, $\xi_{G,\varphi,k}$ and $\xi_{G,\vartheta,k}$). Thus, the reflecting phase of the $k$-th RISS is
\begin{align}
	\boldsymbol{\Theta}^e_k=\text{diag}\left\{\left(\boldsymbol{\alpha}(\vartheta^e_{h,k}, \varphi^e_{h,k})\circ\boldsymbol{\alpha}(\vartheta^e_{G,k}, \varphi^e_{G,k})\right)^\dagger\right\}.
\end{align}
\begin{lemma}\label{lemma:lemma2}
The average energy at the MU can be expressed as Eq. \eqref{eqn:mean_final} at the top of the next page, where $\sigma^2_{\varphi,j}=\sigma^2_{h,\varphi,j}+\sigma^2_{G,\varphi,j}$, $\sigma^2_{\vartheta,j}=\sigma^2_{h,\vartheta,j}+\sigma^2_{G,\vartheta,j}$, $\zeta_k=\varrho_{B2R,k}\varrho_{R2U,k}\sqrt{\eta_k}\sqrt{M}$.
\end{lemma}
\begin{figure*}[ht]
	\begin{equation}
	\begin{aligned}
		\mathbb{E}\{\left|y_{MU}\right|^2\}=\sum_{k=1}^{N_{R2U}}\zeta_k^2\left(\sum_{i=0}^{N_x-1} \sum_{j=0}^{N_x-1} e^{-\frac{(i-j)^2\sigma_{\varphi, j}^2}{2}}\right)& \left(\sum_{i=0}^{N_y-1} \sum_{j=0}^{N_y-1} e^{-\frac{(i-j)^2\sigma_{\vartheta, j}^2}{2}}\right)\\
		+2\sum_{i=1}^{N_{R2U}}\sum_{j=i+1}^{N_{R2U}}&\zeta_i\zeta_j\sum_{n_x=0}^{N_x-1}e^{-\frac{n_x^2\sigma^2_{\varphi,i}}{2}}\sum_{n_x=0}^{N_x-1}e^{-\frac{n_x^2\sigma^2_{\varphi,j}}{2}}\sum_{n_y=0}^{N_y-1}e^{-\frac{n_y^2\sigma^2_{\vartheta,i}}{2}}\sum_{n_y=0}^{N_y-1}e^{-\frac{n_y^2\sigma^2_{\vartheta,j}}{2}}.\label{eqn:mean_final}
	\end{aligned}
\end{equation}
\hrulefill
\end{figure*}

\begin{IEEEproof}
	Please refer to Appendix \ref{app:B} for detailed proof.
\end{IEEEproof}

Note that Eq. \eqref{eqn:mean_final} represents the average received energy while accounting for DOA errors, effectively demonstrating the effectiveness of the sensing-assisted communication scheme. By utilizing the conclusion derived from Eq. \eqref{eqn:mean_final}, we can employ the Jensen's inequality to obtain an upper bound for the ergodic spectral efficiency (ESE) as
\begin{align}
	C\leq\log_2\left(1+\frac{\mathbb{E}\left\{\left|y_{MU}\right|^2\right\}}{\sigma_0^2}\right).
\end{align}

\section{Numerical Results}
This section presents numerical results for evaluating the performance of the proposed scheme and analysis. The BS is equipped with a total of $M=64$ antennas, while every RISS comprises $N = 625$ passive elements. The influence of the number of active elements, $N_a$, is not directly accounted for but is implicitly included by considering its effect on estimation inaccuracies. In this system setup, we have a single MU equipped with one antenna. Furthermore, the RCS is assigned as $\varsigma=100$ $\text{m}^2$ \cite[Table 2.2]{skolnik1980introduction}, the noise power is set to $\sigma_0^2=-94$ dBm, and the signal frequency is 3.5 GHz. The BS, RISSs, and MU are positioned at $(0, 0, 15)$, $({\{x_{R,n}\}_{n=1}^{N_{R2U}}}, 50, 15)$ and $(10, 10, 0)$ respectively, unless specified otherwise.

Fig. \ref{fig:exp1sensingarea} illustrates the sensing detectable range as a function of varying $N_{R2U}$ and transmit power $P$. Here, $\mathcal{A}_{\Sigma} = \sum_{k \in N_{R2U}} \mathcal{A}_k(d_{R2U,k})$ represents the directly sum of the detectable ranges of individual RISSs, where $\mathcal{A}_k(d_{R2U,k})$ denotes the volume of a hemisphere with radius $d_{R2U,k}$. As $N_{R2U}$ increases, the overall detectable range expands. This is expected since despite the range per RISS decreases, the number of RISSs increases. However, when $N_{R2U}$ reaches a certain threshold, the total detectable range begins to reduce due to overlapping among the RISSs. Moreover, a comparison between $\mathcal{A}_{\Sigma}$ and $\mathcal{A}_{\cup}$ reveals earlier differences with power increases, indicating premature overlap among the RISSs' detectable ranges. This is because, as transmission power increases, the power allocated to each RISS also increases, resulting in larger detectable ranges for individual RISS. However, with RISS deployment fixed, the larger $d_{R2U,k}$ leads to more overlap among the coverage areas of the RISSs, thereby reducing the effective total detectable range.

Fig. \ref{fig:exp2spectral_effi} shows the variation in spectral efficiency as the MU moves along the x-axis from 0 m to 150 m. For $N_{R2U} \geq 3$, the spectral efficiency initially rises and then falls, indicating an optimal communication spot during the MU's movement. When the MU is close to the BS (e.g., $x_U = 0$), distant RISSs contribute less, and power is allocated primarily to nearby RISSs. As the MU moves, more RISSs are powered, leading to an enhancement in spectral efficiency. However, increased path loss eventually causes a decline in spectral efficiency. Furthermore, as $N_{R2U}$ increases, the peak spectral efficiency point shifts further from the BS.

\begin{figure}[!t]
	\centering
	\subfloat[The average energy $\mathbb{E}\{\left|y_{MU}\right|^2\}$.]{\includegraphics[width=0.7\linewidth]{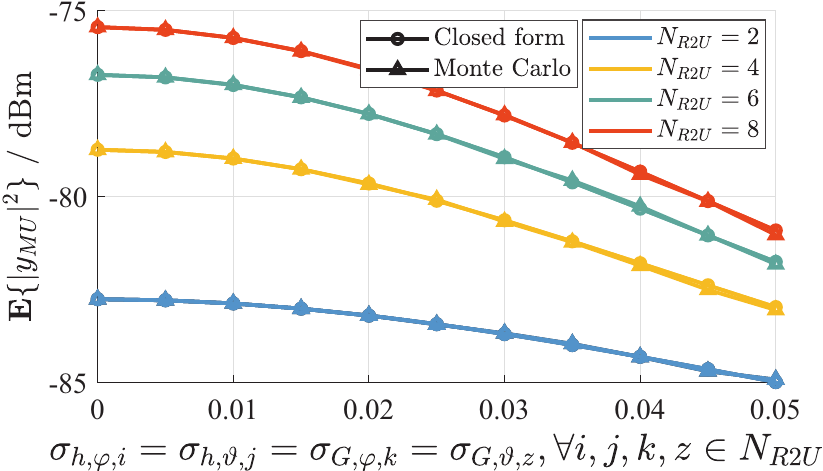}
	\label{fig:receivedenergy}}\\
	\setlength{\abovecaptionskip}{0pt}
	\setlength{\belowcaptionskip}{0pt}
	\vspace{-0.1cm}
	\subfloat[The ergodic spectral efficiency.]{\includegraphics[width=0.7\linewidth]{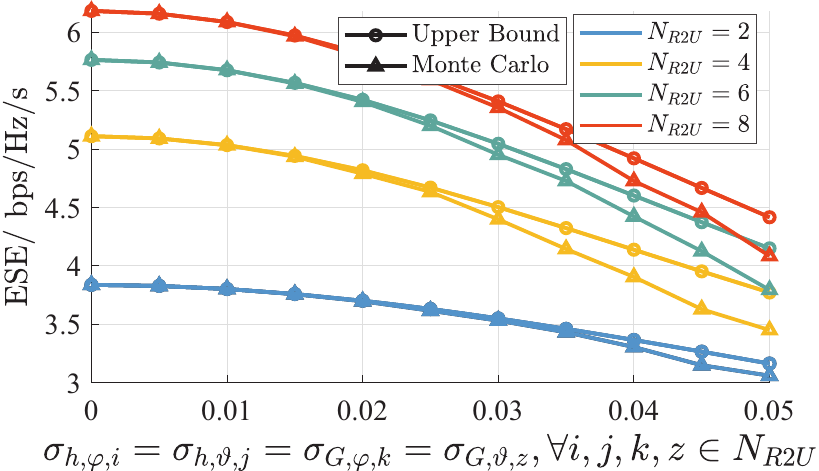}
	\label{fig:speceff}}
	\setlength{\abovecaptionskip}{0pt}
	\setlength{\belowcaptionskip}{0pt}
	\caption{Performance analysis in the existence of sensing errors.}
	\label{fig:exp3sensingassisted}
	\vspace{-0.4cm}
\end{figure}

Fig. \ref{fig:exp3sensingassisted} demonstrates the energy received by the MU at the location of $(50,10,0)$ and the ergodic spectral efficiency in the presence of a sensing error. The transmission power is fixed at $1$ mW. The results underscore that both the received energy and ergodic spectral efficiency decrease drastically as the sensing error rises. Fig. \ref{fig:receivedenergy} validates the precision of the closed-form expression derived in Eq. \eqref{eqn:mean_final}. However, due to the application of Jensen's inequalities, the actual ergodic spectral efficiency deviates from the upper limit as the error increases, as displayed in Fig. \ref{fig:speceff}. However, within the typical range of errors encountered in practical systems, i.e, $\sigma_{h,\varphi,i}=\sigma_{h,\vartheta,j}=\sigma_{G,\varphi,k}=\sigma_{G,\vartheta,z}\leq 0.02\pi, \forall i,j,k,z\in N_{R2U}$, these discrepancies remain negligible. Furthermore, it can be observed that a greater value of $N_{R2U}$ results in a considerable enhancement in the overall system performance.
\section{Conclusion}
In this letter, we investigated the design of multi-RISS-assisted schemes for integrated sensing and communication. We considered various aspects, including RISS deployment, power allocation, and sensing-assisted communication performance analysis. Our numerical results indicate that while the sensing scheme performs optimally with a limited number of RISSs due to overlapping detectable ranges, the communication scheme benefits from a larger number of RISSs. Additionally, we identified an optimal service location for communication. Rather than relying on traditional CSI estimation, future work could explore using sensing information as a substitute for CSI, which may further enhance the performance and applicability of integrated sensing and communication systems, particularly in scenarios where conventional CSI acquisition is challenging or impractical.

{\appendices
\section{Proof of Lemma \ref{lemma:lemma1}}\label{app:A}
As the leakage signals eliminated, we only need to focus on maximizing the $\mathbf{h}_k^T\boldsymbol{\Theta}_k\mathbf{G}_k\mathbf{w}_k,k\in N_{R2U}$, where
\begin{align}
	\mathbf{h}_k^T\boldsymbol{\Theta}_k\mathbf{G}_k\mathbf{w}_k=&\boldsymbol{\alpha}(\vartheta_{h,k}, \varphi_{h,k})^T\mathbf{\Theta}_k\boldsymbol{\alpha}(\vartheta_{G,k}, \varphi_{G,k})\nonumber\\
	&\qquad \qquad \qquad \qquad \times\boldsymbol{\beta}^T(\varpi_{G,k})\mathbf{w}_k.
\end{align}
Subsequently, the optimization process involves independently maximizing $\boldsymbol{\alpha}(\vartheta_{h,k}, \varphi_{h,k})^T\mathbf{\Theta}k\boldsymbol{\alpha}(\vartheta{G,k}, \varphi_{G,k})$ and $\boldsymbol{\beta}^T(\varpi_{G,k})\mathbf{w}_k$, resulting in the desired outcomes.

To eliminate the leakage signals, we have
\begin{align}
	\boldsymbol{\beta}^T(\varpi_{G,k})\boldsymbol{\beta}^\dagger(\varpi_{G,i})=\sum_{m=1}^{M}e^{\mathbbm{i}\pi(m-1)\left(\sin(\varpi_{G,k})-\sin(\varpi_{G,i})\right)}=0,
\end{align}
which means that
\begin{align}
	\left|\sin(\varpi_{G,k})-\sin(\varpi_{G,i})\right|=\frac{2n}{M},n=1,2,\cdots.
\end{align}

Thus, the proof is completed.

\section{Proof of Lemma \ref{lemma:lemma2}}\label{app:B}
We have
\begin{align}
	&\boldsymbol{\alpha}(\vartheta_{h,k}, \varphi_{h,k})^T\mathbf{\Theta}^e_k\boldsymbol{\alpha}(\vartheta_{G,k}, \varphi_{G,k})\nonumber\\
	=&\mathbf{1}_{1\times N}\boldsymbol{\alpha}(\xi_{h,\vartheta,k}+\xi_{G,\vartheta,k},\xi_{h,\varphi,k}+\xi_{G,\varphi,k}),
\end{align}
since $\mathbf{A}^T\mathbf{\Theta}\mathbf{B}=\text{diag}\{\mathbf{\Theta}\}^T\left(\mathbf{A}\circ\mathbf{B}\right)$, $\boldsymbol{\alpha}(\vartheta, \varphi) = \boldsymbol{\alpha}_x(\varphi)\otimes \boldsymbol{\alpha}_y(\vartheta)$, $(\mathbf{A}\otimes\mathbf{B})^H=\mathbf{A}^H\otimes\mathbf{B}^H$,  $(\mathbf{A}\otimes\mathbf{B})\circ(\mathbf{C}\otimes\mathbf{D})=(\mathbf{A}\circ\mathbf{C})\otimes(\mathbf{B}\circ\mathbf{D})$ and $(\mathbf{A}\otimes\mathbf{B})(\mathbf{C}\otimes\mathbf{D})=(\mathbf{A}\mathbf{C})\otimes(\mathbf{B}\mathbf{D})$.

Thus, we can obtain the expression of Eq. \eqref{eqn:twopart} at the top of this page,
\begin{figure*}[ht]
	\begin{equation}
	\begin{aligned}
		&\mathbb{E}\{\left|y_{MU}\right|^2\}=
		\mathbb{E}\left\{\left|\sum_{k=1}^{N_{R2U}}\zeta_k\mathbf{1}_{1\times N}\boldsymbol{\alpha}(\xi_{h,\vartheta,k}+\xi_{G,\vartheta,k},\xi_{h,\varphi,k}+\xi_{G,\varphi,k})\right|^2\right\}\\
		=&\underbrace{\sum_{k=1}^{N_{R2U}}\mathbb{E}\left\{\left|\zeta_k\mathbf{1}_{1\times N}\boldsymbol{\alpha}(\xi_{h,\vartheta,k}+\xi_{G,\vartheta,k},\xi_{h,\varphi,k}+\xi_{G,\varphi,k})\right|^2\right\}}_{\text{Part 1}}\\
		&+\underbrace{\mathbb{E}\left\{2\Re\left(\sum_{i=1}^{N_{R2U}}\sum_{j=i+1}^{N_{R2U}}\zeta_i\zeta_j\mathbf{1}_{1\times N}\boldsymbol{\alpha}(\xi_{h,\vartheta,i}+\xi_{G,\vartheta,i},\xi_{h,\varphi,i}+\xi_{G,\varphi,i})\boldsymbol{\alpha}^H(\xi_{h,\vartheta,j}+\xi_{G,\vartheta,j},\xi_{h,\varphi,j}+\xi_{G,\varphi,j})\mathbf{1}_{N\times 1}\right),\label{eqn:twopart}\right\}}_{\text{Part 2}}
	\end{aligned}
\end{equation}
\hrulefill
\end{figure*}
where $\zeta_k=\varrho_{B2R,k}\varrho_{R2U,k}\sqrt{\eta_k}\sqrt{M}$.

The first component of Eq. \eqref{eqn:twopart} can be expressed as 
	\begin{align}
		&\text{Part 1}\nonumber\\
		&=\sum_{k=1}^{N_{I2U}}\mathbb{E}\left\{\left|\zeta_k\mathbf{1}_{1\times N}\boldsymbol{\alpha}(\xi_{h,\vartheta,k}+\xi_{G,\vartheta,k},\xi_{h,\varphi,k}+\xi_{G,\varphi,k})\right|^2\right\},
	\end{align}
	where the $k$-th component of the summation can be rewritten as 
	\begin{align}
		&\mathbb{E}\left\{\left|\zeta_k\mathbf{1}_{1\times N}\boldsymbol{\alpha}(\xi_{h,\vartheta,k}+\xi_{G,\vartheta,k},\xi_{h,\varphi,k}+\xi_{G,\varphi,k})\right|^2\right\}\nonumber\\
		\overset{(a)}{=}&\zeta_k^2\mathbb{E}\left\{\left|\sum_{i=0}^{N_x-1}e^{\mathbbm{i}i(\xi_{h,\vartheta,k}+\xi_{G,\vartheta,k})}\right|^2\right\}\nonumber\\
		&\qquad\qquad\qquad\qquad \times\mathbb{E}\left\{\left|\sum_{j=0}^{N_y-1}e^{\mathbbm{i}j(\xi_{h,\varphi,k}+\xi_{G,\varphi,k})}\right|^2\right\}\nonumber\\
		\overset{(b)}{=}&\zeta_k^2\mathbb{E}\left\{\sum_{i=0}^{N_x-1}\sum_{m=0}^{N_x-1}\cos\left((i-m)(\xi_{h,\vartheta,k}+\xi_{G,\vartheta,k})\right)\right\}\nonumber\\
		&\qquad \times\mathbb{E}\left\{\sum_{j=0}^{N_y-1}\sum_{n=0}^{N_y-1}\cos\left((j-n)(\xi_{h,\varphi,k}+\xi_{G,\varphi,k})\right)\right\}\nonumber\\
		\overset{(c)}{=}&\sum_{i=0}^{N_x-1}\sum_{m=0}^{N_x-1}\sum_{l =0}^\infty\frac{(-1)^{l}}{(2l)!}\mathbb{E}\left\{\left((i-m)(\xi_{h,\vartheta,k}+\xi_{G,\vartheta,k})\right)^{2l}\right\}\nonumber\\
		&\times\sum_{j=0}^{N_y-1}\sum_{n=0}^{N_y-1}\sum_{l =0}^\infty\frac{(-1)^{l}}{(2l)!}\mathbb{E}\left\{\left((j-n)(\xi_{h,\varphi,k}+\xi_{G,\varphi,k})\right)^{2l}\right\}.\label{eqn:taylorseries}
	\end{align}
	Where $(a)$ comes from that $\boldsymbol{\alpha}(\xi_{h,\vartheta,k}+\xi_{G,\vartheta,k},\xi_{h,\varphi,k}+\xi_{G,\varphi,k})=\boldsymbol{\alpha}_x(\xi_{h,\vartheta,k}+\xi_{G,\vartheta,k})\otimes\boldsymbol{\alpha}_y(\xi_{h,\varphi,k}+\xi_{G,\varphi,k})$, and $\boldsymbol{\alpha}_x(\xi_{h,\vartheta,k}+\xi_{G,\vartheta,k})=[1, \cdots,e^{\mathbbm{i}(N_x-1)(\xi_{h,\vartheta,k}+\xi_{G,\vartheta,k})}]^T$. $(b)$ is due to Euler's formula and prosthaphaeresis, and $(c)$ comes from the Taylor series. We make $\Psi=\left((i-m)(\xi_{h,\vartheta,k}+\xi_{G,\vartheta,k})\right)^{2l}$, and since $\xi_{h,\vartheta,k}+\xi_{G,\vartheta,k}\sim\mathcal{N}(0,\sigma^2_{\vartheta,k})$, where $\sigma^2_{\vartheta,k} = \sigma^2_{h,\vartheta,k}+\sigma^2_{G,\vartheta,k}$, we have the probability distribution function (PDF) of $\Psi$ is
	\begin{align}
		f_{\Psi}(\psi)=f_{\xi_{h,\vartheta,k}+\xi_{G,\vartheta,k}}(\psi^{\frac{1}{2l}})\frac{1}{l}\psi^{\frac{1}{2l}-1}=\frac{\psi^{\frac{1}{2l}-1}}{\sqrt{2\pi} l\sigma_{\vartheta,k}}e^{-\frac{\psi^{\frac{1}{l}}}{2\sigma_{\vartheta,k}^2}},
	\end{align}
	where $f_{\Psi}(\cdot)$ and $f_{\xi_{h,\vartheta,k}+\xi_{G,\vartheta,k}}(\cdot)$ represent the PDF of variable $\Psi$ and ${\xi_{h,\vartheta,k}+\xi_{G,\vartheta,k}}$, respectively. Thus we hare
	\begin{align}
		&\mathbb{E}\left\{\left((i-m)({\xi_{h,\vartheta,k}+\xi_{G,\vartheta,k}})\right)^{2l}\right\}\nonumber\\
		=&\int_0^\infty \psi f_{\Psi}(\psi)\text{d} \psi=\int_0^\infty\frac{\psi^{\frac{1}{2l}}}{\sqrt{2\pi} l(i-m)\sigma_{\vartheta,k}}e^{-\frac{\psi^{\frac{1}{l}}}{2(i-m)^2\sigma_{\vartheta,k}^2}}\text{d}\psi\nonumber\\
		=&\frac{(2l)!}{l!2^l}\left((i-m)\sigma_{\vartheta,k}\right)^{2l}.\label{eqn:expeofvarXI}
	\end{align}
	Take Eq. \eqref{eqn:expeofvarXI} into Eq. \eqref{eqn:taylorseries}, yielding
	\begin{align}
		&\sum_{i=0}^{N_x-1}\sum_{m=0}^{N_x-1}\sum_{l=0}^\infty\frac{(-1)^{l}}{(2l)!}\mathbb{E}\left\{\left((i-m)(\xi_{h,\vartheta,k}+\xi_{G,\vartheta,k})\right)^{2l}\right\}\nonumber\\
		=&\sum_{i=0}^{N_x-1}\sum_{m=0}^{N_x-1}\sum_{l=0}^\infty\frac{(-1)^{l}}{(2l)!}\frac{(2l)!}{l!2^l}\left((i-k)\sigma_{\vartheta,k}\right)^{2n}\nonumber\\
		=&\sum_{i=0}^{N_x-1}\sum_{m=0}^{N_x-1}e^{-\frac{(i-m)^2\sigma_{\vartheta,k}^2}{2}}.
	\end{align}
	And finally we can obtain
	\begin{align}
	&\mathbb{E}\left\{\left|\zeta_k\mathbf{1}_{1\times N}\boldsymbol{\alpha}(\xi_{h,\vartheta,k}+\xi_{G,\vartheta,k},\xi_{h,\varphi,k}+\xi_{G,\varphi,k})\right|^2\right\}\nonumber\\
	=&\sum_{i=0}^{N_x-1}\sum_{m=0}^{N_x-1}\sum_{l =0}^\infty\frac{(-1)^{l}}{(2l)!}\mathbb{E}\left\{\left((i-m)(\xi_{h,\vartheta,k}+\xi_{G,\vartheta,k})\right)^{2l}\right\}\nonumber\\
	&\times\sum_{j=0}^{N_y-1}\sum_{n=0}^{N_y-1}\sum_{l =0}^\infty\frac{(-1)^{l}}{(2l)!}\mathbb{E}\left\{\left((j-n)(\xi_{h,\varphi,k}+\xi_{G,\varphi,k})\right)^{2l}\right\}\nonumber\\
	=&\left(\sum_{i=0}^{N_x-1}\sum_{m=0}^{N_x-1}e^{-\frac{(i-m)^2\sigma_{\vartheta,k}^2}{2}}\right)\left(\sum_{j=0}^{N_y-1}\sum_{n=0}^{N_y-1}e^{-\frac{(j-n)^2\sigma_{\varphi,k}^2}{2}}\right).
	\end{align}
	Similarly, the second component of Eq. (25) is expressed as
	\begin{align}
		\text{Part 2}
		=&\mathbb{E}\left\{2\Re\left(\sum_{i=1}^{N_{R2U}}\sum_{j=i+1}^{N_{R2U}}\zeta_i\zeta_j\sum_{n_x=0}^{N_x-1}e^{\mathbbm{i}n_x\xi_{\varphi,i}}\right.\right.\nonumber\\
		&\left.\left.\times\sum_{n_y=0}^{N_y-1}e^{\mathbbm{i}n_y\xi_{\vartheta,i}}\sum_{n_x=0}^{N_x-1}e^{-\mathbbm{i}n_x\xi_{\varphi,j}}\sum_{n_y=0}^{N_y-1}e^{-\mathbbm{i}n_y\xi_{\vartheta,j}}\right)\right\}\nonumber\\
		=&2\sum_{i=1}^{N_{R2U}}\sum_{j=i+1}^{N_{R2U}}\zeta_i\zeta_j\sum_{n_x=0}^{N_x-1}e^{-\frac{n_x^2\sigma^2_{\varphi,i}}{2}}\sum_{n_x=0}^{N_x-1}e^{-\frac{n_x^2\sigma^2_{\varphi,j}}{2}}\nonumber\\
		&\qquad\qquad\quad\times\sum_{n_y=0}^{N_y-1}e^{-\frac{n_y^2\sigma^2_{\vartheta,i}}{2}}\sum_{n_y=0}^{N_y-1}e^{-\frac{n_y^2\sigma^2_{\vartheta,j}}{2}}.\label{eqn:secondpart}
	\end{align}
	
	Thus, the proof is completed.

}
    \bibliography{Reference}
	\bibliographystyle{IEEEtran}

\end{document}